\documentclass{llncs}

\usepackage{color}
\usepackage{tcolorbox}
\usepackage{gensymb}
\usepackage{amsmath}
\usepackage{makeidx}  
\usepackage{amssymb}
\usepackage{graphicx}
\usepackage{mathrsfs}
\usepackage{nd3}
\usepackage{pgf}
\usepackage{multirow,array}
\usepackage{mathpartir}
\usepackage{float}
\usepackage{tikz}
\usetikzlibrary{arrows,automata,positioning}
\usepackage{stmaryrd}
\usepackage{listings}
\usepackage{fancyvrb}
\usepackage{alltt}
\definecolor{dkviolet}{rgb}{0.58,0,0.83}
\definecolor{ltblue}{rgb}{0, 0, 0.55}
\definecolor{dkgreen}{rgb}{0.0, 0.2, 0.13}
\definecolor{dkblue}{rgb}{0,0,1}
\definecolor{dkred}{rgb}{0.55,0,0}
\lstdefinelanguage{Coq}{ 
	mathescape=true,
	texcl=false, 
	morekeywords=[1]{Function, Section, Module, End, Require, Import, Export,
		Variable, Variables, Parameter, Parameters, Axiom, Hypothesis,
		Hypotheses, Notation, Local, Tactic, Reserved, Scope, Open, Close,
		Bind, Delimit, Definition, Let, Ltac, Fixpoint, CoFixpoint, Add,
		Morphism, Relation, Implicit, Arguments, Unset, Contextual,
		Strict, Prenex, Implicits, Inductive, CoInductive, Record,
		Structure, Canonical, Coercion, Context, Class, Global, Instance,
		Program, Infix, Theorem, Lemma, Corollary, Proposition, Fact,
		Remark, Example, Proof, Goal, Save, Qed, Defined, Hint, Resolve,
		Rewrite, View, Search, Show, Print, Printing, All, Eval, Check,
		Projections, inside, outside, Def},
	morekeywords=[2]{forall, exists, exists2, fun, fix, cofix, struct,
		match, with, end, as, in, return, let, if, is, then, else, for, of,
		nosimpl, when},
	morekeywords=[3]{Type, Prop, Set, true, false, option},
	morekeywords=[4]{pose, set, move, case, elim, apply, clear, hnf,
		intro, intros, generalize, rename, pattern, after, destruct,
		induction, using, refine, inversion, injection, rewrite, congr,
		unlock, compute, ring, field, fourier, replace, fold, unfold,
		change, cutrewrite, simpl, have, suff, wlog, suffices, without,
		loss, nat_norm, assert, cut, trivial, revert, bool_congr, nat_congr,
		symmetry, transitivity, auto, split, left, right, autorewrite},
	morekeywords=[5]{by, done, exact, reflexivity, tauto, romega, omega,
		assumption, solve, contradiction, discriminate},
	morekeywords=[6]{do, last, first, try, idtac, repeat},
	morecomment=[s]{(*}{*)},
	showstringspaces=false,
	morestring=[b]",
	morestring=[d],
	tabsize=3,
	extendedchars=false,
	sensitive=true,
	breaklines=false,
	basicstyle=\small,
	captionpos=b,
	columns=[l]flexible,
	identifierstyle={\ttfamily\color{black}},
	keywordstyle=[1]{\ttfamily\color{dkviolet}},
	keywordstyle=[2]{\ttfamily\color{dkgreen}},
	keywordstyle=[3]{\ttfamily\color{ltblue}},
	keywordstyle=[4]{\ttfamily\color{dkblue}},
	keywordstyle=[5]{\ttfamily\color{dkred}},
	stringstyle=\ttfamily,
	commentstyle={\ttfamily\color{dkgreen}},
	literate=
	{\\forall}{{\color{dkgreen}{$\forall\;$}}}1
	{\\exists}{{$\exists\;$}}1
	{<-}{{$\leftarrow\;$}}1
	{=>}{{$\Rightarrow\;$}}1
	{==>}{{$\Longrightarrow\;$}}1
	{->}{{$\rightarrow\;$}}1
	{<->}{{$\leftrightarrow\;$}}1
	{<==}{{$\leq\;$}}1
	{\#}{{$^\star$}}1 
	{\\o}{{$\circ\;$}}1 
	{\@}{{$\cdot$}}1 
	{\/\\}{{$\wedge\;$}}1
	{\\\/}{{$\vee\;$}}1
	{~}{{\ }}1
	{\@\@}{{$@$}}1
	{\\mapsto}{{$\mapsto\;$}}1
	{\\hline}{{\rule{\linewidth}{0.5pt}}}1
}[keywords,comments,strings]

\lstnewenvironment{coq}{\lstset{language=Coq}}{}

\newcommand{\Rel}[1]{R_{#1}}

\newcommand{\Kns}[1]{\mathbf{K}_{\mathbf{#1}}\,} 
 
\newcommand{\Bels}[1]{\mathbf{B}_{\mathbf{#1}}\,} 
 
\newcommand{\Poss}[1]  {\langle \mathbf{K}_{#1} \rangle\,}

\newcommand{\BPoss}[1]{\langle \mathbf{B}_{#1} \rangle\,}

\newcommand{\Pal}[1]{\mathbf{[{#1}]}}
\newcommand{\PalPos}[2]{\mathbf{\langle #1, #2 \rangle}\,}

\newcommand{\SPalPos}[2]{\mathbf{\langle #1, #2,\mathbf{S} \rangle}\,}

\newcommand{\bnf}{\ |\ }

\newcommand{\iiff}{\textsf{\  iff \ }}
\newcommand {\aand}{ \textsf{\ and \ } }
\newcommand {\timplies} { \textsf{\ implies \ }}
\newcommand {\tland}{{\ \wedge \ }}
\newcommand {\tlor}{{\ \vee \ }}
\newcommand {\tlnot}{{\neg}}
\newcommand {\iimplies}{{\ \Rightarrow \ }}
\makeatletter
\def\BState{\State\hskip-\ALG@thistlm}
\makeatother
\newcommand{\pre}{\mathit{pre}}
\newcommand{\lpp} {\mathit{p}}

\newcommand{\laa} {a}
\newcommand{\lbb} {b}

\begin{document}
%
%
\pagestyle{headings}  
%

%
%
%
\title{Reasoning About Safety-Critical Information Flow Between Pilot and Computer}
\author{Seth Ahrenbach}
%
%
%
\institute{University of Missouri, Columbia MO 65201, USA,\\
\email{SJK7v7@mail.missouri.edu}}
\mainmatter
\maketitle              
\begin{abstract}
	This paper presents research results that develop a dynamic logic for reasoning about safety-critical information flow among humans and computers. The logic advances previous efforts to develop logics of agent knowledge, which make assumptions that are too strong for realistic human agents. We introduce Dynamic Agent Safety Logic (DASL), based on Dynamic Epistemic Logic (DEL), with extensions to account for safe actions, belief, and the logical relationships among knowledge, belief, and safe action. With this logic we can infer which safety-critical information a pilot is missing when executing an unsafe action. We apply the logic to the Air France 447 incident as a case study and provide a mechanization of the case study in the Coq proof assistant.
\end{abstract}
\section{Introduction}

A common theme for aviation mishaps attributed to human error is for a pilot to become overwhelmed by data, lose situational awareness, and provide unsafe inputs to the flight controls. As yet, little work has been done to leverage the power of formal methods to address this problem. This paper remedies that by defining a dynamic logic of belief, knowledge, and safe action. We use the logic to create an axiomatic model of agency suitable for reasoning about safety-critical information flow among pilots and the flight computer. We mechanize this model in the Coq Proof Assistant and apply it to the Air France 447 incident as a case study.\footnote{This is a pre-print of an article published in \emph{9th International Symposium, NASA Formal Methods 17}. The final authenticated version is available online at: https://doi.org/10.1007/978-3-319-57288-8\_25.}\footnote{Code: https://github.com/sethahrenbach/DASL/blob/master/DASL.v} 

The research contributions of this paper include the development of a dynamic logic that is suitable for reasoning about safety-critical information flow. The dynamic logic is extended beyond most dynamic logics' treatment of action in that it treats both \emph{mere} action and \emph{safe} action, and captures the relationship between the two. The subsequent application and mechanization in Coq explore novel uses of formal methods in aviation safety, beyond mere verification of system component correctness. They introduce the idea of formally analyzing the human component of the safety-critical systems.

Dynamic Logic is a type of modal logic used for reasoning about state transition diagrams of programs~\cite{DynLog,modal}. A diagram consists of nodes and edges, representing states of the system and labeled transitions between them, respectively. It is distinguished from other logics by the fact that truth is dynamic, rather than static, in its semantics. Thus, it is capable of representing the way actions change the truth of propositions. It serves as a foundation for a variety of logics similarly concerned with changes in some aspect of the truth as a result of actions. This family of logics has been described as logical dynamics, and includes Public Announcement Logic (PAL) and Dynamic Epistemic Logic (DEL)~\cite{VB_LDII}.

Logical dynamics allows researchers to model information flow, rationality, and action in multi-agent systems~\cite{VB_LDII}. In Ahrenbach and Goodloe~\cite{AhrenbachGoodloe}, the authors develop a static modal logic for knowledge, belief, and safety to analyze a family of aviation mishaps involving a type of reasoning error suffered by a single pilot. This paper extends that work by employing dynamic methodologies from logical dynamics to the analysis of mishaps. The use of a dynamic logic rather than a static logic connects safety-critical information and actions in a more natural way, and allows for easier inference from action to information. The application of these methods advances the discipline of logical dynamics by employing them in the real world, beyond toy examples and logic puzzles, and likewise improves the discipline of aviation safety by introducing a formal method suitable for analyzing safety-critical information flow between pilots and machine.

Recent work at the intersection of game theory and logical dynamics focuses on information flow during games. Van Ditmarsch identifies a class of games called \emph{knowledge games}, in which players have diverging information\cite{ditmarsch}. This slightly relaxes the assumption of classical game theory that players have common knowledge about each other's perfect information. This invites logicians to study the information conveyed by the fact that an action is executed. For example, if agent $1$ asks agent $2$ the question, ``$p$?", the information conveyed is that $1$ does not know whether $p$, believes that $2$ knows whether $p$, and after the action occurs, this information becomes publicly known. Many actions convey such information, beyond mere speech acts. For example, when a pilot provides flight control inputs, her action conveys information about what she believes about the aircraft's state, namely that it is in a state that safely permits those inputs. Anyone observing her inputs, like the first officer or the flight computer, can make such inferences about her mental picture based on her actions.

This paper proceeds as follows. In Section~\ref{DASL} we define the formal model, which consists of a set of axioms in a dynamic modal logic for reasoning about pilot knowledge, belief, and safety. Section~\ref{mech} mechanizes the model in the Coq Proof Assistant and applies it to case studies, illustrating the logic's use as a formal method for aviation safety. We offer a brief discussion of future work in Section~\ref{futurework} and conclude in Section~\ref{conc}.

\newpage
\section{Dynamic Agent Safety Logic}~\label{DASL}
The logic for reasoning about information flow in knowledge games is called Dynamic Epistemic Logic (DEL). As its name suggests, it combines elements of epistemic logic and dynamic logic. Epistemic logic is the static logic for reasoning about knowledge, and dynamic logic is used to reason about actions. In dynamic logic semantics, nodes are states of the system or the world, and relations on nodes are transitions via programs or actions from node to node. If we think of each node in dynamic logic as being a model of epistemic logic, then actions become relations on models, representing transitions from one multi-agent epistemic model to another. For example, if we have a static epistemic model $M1$ representing the knowledge states of agents $1$ and $2$ at a moment, then the action $``p?"$ is a relation between $M1$ and $M2$, a new static epistemic model of $1$'s and $2$'s knowledge after the question is asked. All of this is captured by DEL.





 We are concerned with an additional element: the \emph{safety} status of an action, and an agent's knowledge and belief about that. To capture this, we extend DEL and call the new logic Dynamic Agent Safety Logic (DASL). The remainder of this section presents DASL's syntax, semantics, and proves its soundness. 
\subsection{Syntax and Semantics}
The Dynamic Agent Safety Logic (DASL) used in this paper has the following syntax.

\begin{tcolorbox}
$$ \varphi \ ::=\   \lpp  \bnf \tlnot \varphi \bnf \varphi \tland \varphi  \bnf \Kns{i} \varphi \bnf \Bels{i}\varphi \bnf \Pal{i, (A,\laa)}\varphi \bnf \Pal{i, (A,\laa),S}\varphi,$$
\end{tcolorbox}
where $\lpp \in AtProp$ is an atomic proposition, $\mathbf{i}$ refers to $i \in Agents$, $\mathbf{\laa}$ is the name of an action, called an action token, belong to a set of such tokens, $Actions$, and $\mathbf{A}$ refers to an action structure. The knowledge operator $\Kns{i}$ indicates that ``agent \emph{i} knows that ..." Similarly, the operator for belief, $\Bels{i}$ can be read, ``agent \emph{i} believes that..." The notion of action tokens and structures will be defined in the semantics. The operators $\Pal{i,(A,\laa)}$ and $\Pal{i,(A,\laa),S}$ are the dynamic operators for agent $i$ executing action token $\laa$ from action structure $A$ in the former case, and doing so safely in the latter case. Note that the $\mathbf{S}$ in $\Pal{i,(A,\laa),S}$ stands for `safety', and is not a variable, whereas the $\mathbf{i,(A,\laa)}$ are variables for agents, action structures, and action tokens, respectively. One can read the action operators as ``after $i$ executes $\laa$ from $A$, $\varphi$ holds.' We define the dual modal operators $\Poss{i}$, $\BPoss{i}$, $\PalPos{i}{(A,\laa)}$, and $\SPalPos{i}{(A,\laa)}$ in the usual way. 

The semantics of DASL involve two structures that are defined simultaneously, one for epistemic models, and one for action structures capturing the transition relation among epistemic models. Additionally, we define numerous helper functions that straddle the division between metalanguage and object language. 

\subsubsection{Kripke Model}
A Kripke model $M\in Model$ is a tuple $\langle W, \{\Rel{k}^i\}, \{\Rel{b}^i\}, w, V \rangle$. It is a set of worlds, sets of epistemic and doxastic relations on worlds for agents, a world denoting the actual world, and a valuation function \emph{V} mapping atomic propositions to the set of worlds satisfying them. Most readers will be somewhat familiar with epistemic logic, the logic for reasoning about knowledge. Doxastic logic is a similar logic for reasoning about belief\cite{Hintikka}.

\subsubsection{Action Structure}
An action structure $A\in ActionStruct$ is a tuple $\langle Actions,\\\{\chi_{k}^i\}, \{\chi_{b}^i\}, \laa \rangle$. It is a set of action tokens, sets of epistemic and doxastic relations on action tokens for agents, and an action token, $\laa$, denoting an actual action token executed. 

An action structure captures the associated subjective events of an action occurring, including how it is observed by various agents, incorporating their uncertainty. The action tokens are the actual objective events that might occur. For example, if I am handed a piece of paper telling me who won the Oscar for Best Actress, and I read it, and you see me read it, then the action structure will include possible tokens in which I read that each nominee has won, and you will consider each of these tokens to be possible. When I read the paper, I consider only one action token to be the one executed. This action structure represents that transition from one epistemic model, in which both of us considers all nominees the potential winner, to an epistemic model in which I know the winner and you still do not know the winner. We can think of the action structure $A$ as the general action ``Agent 1 reads the piece of paper" and the tokens as the specific actions ``Agent 1 reads that nominee \emph{n} has won the award."

\subsubsection{Model Relation}
Just as $\Rel{k}^i$ denotes a relation on worlds, $\llbracket i,(A,\laa) \rrbracket$ denotes a relation on Kripke model-world pairs. It represents the relation that holds between $M,w$ and $M',w'$ when agent $i$ executes action $(A,\laa)$ at $M,w$ and causes the world to transition to $M',w'$.

\subsubsection{Precondition Function}
The Precondition function, $pre :: Actions \mapsto \varphi$, maps an action to the formula capturing the conditions under which the action can occur. For example, if we assume agents tell the truth, then an announcement action has as a precondition that the announced proposition is true, as with regular Public Announcement Logic.


\subsubsection{Postcondition Function}
The Postcondition function, $post :: A \times AtProp \mapsto AtProp$, takes an action structure and an atomic proposition, and maps to the corresponding atomic proposition after the action occurs.
\begin{align*}
post(A,p)= p\ \mbox{if}\ update(M,A,w,\laa,i)\models p,\  \mbox{else}\ \tlnot p.
\end{align*} 

\subsubsection{Update Function}
The Update function, $update :: (Model \times ActionStruct \times W \times Actions \times Agents) \mapsto (Model \times W)$, takes a Kripke model $M$, an action structure $A$, a world from the Kripke model, an action token from the Action structure, and an agent executing the action, and returns a new Kripke model-world pair. It represents the effect actions have on models, and is more complicated than other DEL semantics in that actions can change the facts on the ground in addition to the knowledge and belief relations. It is a partial function that is defined iff a model-world pair satisfies the action's preconditions.
\\

$update(M,A,w,\laa,i) = (M',w')\ where:\\$
$1.\  M = \langle W, \{\Rel{k}^{i}\}, \{\Rel{b}^{i}\}, w, V \rangle$\\
$2.\  A = \langle Actions, \{\chi_{k}^i\}, \{\chi_{b}^i\}, \laa, pre, post \rangle$\\
$3.\  M' = \langle W', \{\Rel{k}'^{i}\}, \{\Rel{b}'^{i}\}, w', V' \rangle$\\
$4.\  W' = \{(w,\laa) | w\in W,\laa \in Actions,\ \aand\ w\models pre(\laa)\}$\\
$5.\  \Rel{k}'^{i} = \{((w,\laa),(v,\lbb))|w\Rel{k}^i v \aand \laa \chi_{k}^i \lbb \}$\\
$6.\  \Rel{b}'^{i} = \{((w,\laa),(v,\lbb))|w\Rel{b}^i v \aand \laa \chi_{b}^i \lbb \}$\\
$7.\  w' = (w,\laa)$\\ 
$8.\  V'(p) = post(A,p)$

\subsubsection{Safety Precondition Function}
The Safety Precondition Function, $pre_s :: Actions \mapsto \varphi$, is a more restrictive function than $pre$. Where $pre$ returns the conditions that dictate whether the action is possible, $pre_s$ returns the conditions that dictate whether the action is safely permissible. This function is the key reason the dynamic approach allows for easy inference from action to safety-critical information.


The logic DASL has the following Kripke semantics.
\begin{tcolorbox}
\begin{align*}
 M,w \models p  &\iiff w \in V(p) \\
 M,w \models \tlnot \varphi  &\iiff M,w \not\models \varphi \\ 
 M,w \models \varphi \tland \psi &\iiff M,w \models \varphi \aand M,w \models \psi \\
 M,w \models \Kns{i}\varphi &\iiff \forall v,\ w\Rel{k}^i v\ \timplies\ M,v \models \varphi \\
 M,w \models \Bels{i}\varphi &\iiff \forall v,\ w\Rel{b}^i v\ \timplies\ M,v \models \varphi \\
M,w \models \Pal{i,(A,\laa)}\varphi &\iiff \forall M',w',\  (M,w) \llbracket i,(A,\laa) \rrbracket (M',w')\ \\&\timplies\ M',w' \models \varphi \\
M,w \models \Pal{i,(A,\laa),S}\varphi &\iiff \forall M',w',\  (M,w) \llbracket i,(A,\laa),S \rrbracket (M',w')\ \\&\timplies\ M',w' \models \varphi 
\end{align*}
\end{tcolorbox}
The definitions of the dynamic modalities make use of a relation between two model-world pairs, which we now define.
\begin{tcolorbox}
	\begin{align*}
	(M,w)\llbracket i,(A,\laa)\rrbracket (M',w') &\iiff M,w \models \pre(\laa) \\&\aand update(M,A,w,\laa,i) = (M',w') \\
(M,w)\llbracket i, (A,\laa),S \rrbracket (M',w') &\iiff M,w \models pre_s(\laa) \\&\aand update(M,A,w,\laa, i) = (M',w') 
    \end{align*}
\end{tcolorbox}

\subsection{Hilbert System}
DASL is axiomatized by the following Hilbert system.\\
All propositional tautologies are axioms.\\
\begin{tcolorbox}$\Kns{i}$ is T (knowledge relation is reflexive)\\
$\Bels{i}$ is KD45 (belief relation is serial, transitive, and Euclidean)\\
EP1: $\Kns{i}\varphi \iimplies \Bels{i}\varphi$ \\
EP2: $\Bels{i}\varphi \iimplies \Bels{i}\Kns{i}\varphi$\\
EP3: $\Bels{i}\varphi \iimplies \Kns{i}\Bels{i}\varphi$\\
SP: $\Pal{i,(A,\laa)}\varphi \iimplies \Pal{i,(A,\laa),S}\varphi$\\
PR: $\PalPos{i}{(A,\laa)}\varphi \iimplies \Bels{i}\SPalPos{i}{(A,\laa)}\varphi$,\\
\end{tcolorbox}
\noindent plus the inference rules Modus Ponens and Necessitation for $\Kns{i}$ and $\Bels{i}$.

Above are the axioms characterizing the logic. Knowledge is weaker here than in most epistemic logics, and belief is standard~\cite{FHMV}. They are related logically by EP(1-3), which hold that knowledge entails belief, belief entails that one believes that one knows, and belief entails than one knows that one believes. Finally, actions and safe actions are logically related by SP and PR, which hold that necessary consequences of \emph{mere} action are also necessary consequences of \emph{safe} actions, and that a pilot can execute an action only if he believes that he is executing a safe action. 



\subsection{Soundness}
\begin{tcolorbox}
\begin{theorem}[Soundness]
	Dynamic Agent Safety Logic is sound for Kripke structures with\\ (1) reflexive $\Rel{k}^i$ relations,\\ (2) serial, transitive, Euclidean $\Rel{b}^i$ relations, \\(3) which are partially ordered $(\Rel{k}^i \circ \Rel{b}^i) \subseteq \Rel{b}^i$, $(\Rel{b}^i \circ \Rel{k}^i) \subseteq \Rel{b}^i$, and $\Rel{b}^i \subseteq \Rel{k}^i$, \\(4) $\llbracket i,(A,\laa),S\rrbracket \subseteq \llbracket i,(A,\laa)\rrbracket$
	and\\ (5) $(\llbracket i,(A,\laa),S\rrbracket \circ \Rel{b}^i) \subseteq \llbracket i, (A,\laa)\rrbracket$.
\end{theorem}
\end{tcolorbox}
$\mathbf{Proof}.$ $(1)\  and\  (2)$ correspond to the axioms that $\Kns{i}$ is a T modality and $\Bels{i}$ is a KD45 modality in the usual way. $(3)$ corresponds to EP1, EP2, and EP3. Axioms AP through SB are reduction axioms. This leaves $(4)$, corresponding to SP, and $(5)$ which corresponds to PR. Here we will prove $(5)$. Let $M$ be a Kripke structure satisfying the five conditions above. Let $A$ be an Action structure with $\laa$ and $i$ as its actual action token and agent. 

We prove $(5)$ via the contrapositive of PR: $\BPoss{i}\Pal{i,(A,\laa),S}\varphi \iimplies \Pal{i, (A,\laa)}\varphi$.
Assume $M,w \models \BPoss{i}\Pal{i.(A,\laa),S}\varphi$. By the semantics of $\BPoss{i}$, there exists a $v$, such that $w\Rel{b}^i v$ and $v \models \Pal{i,(A,\laa),S}\varphi$. From the semantics, it follows that forall $M',v'$, if $(M,v)\llbracket i,(A,\laa),S\rrbracket (M',v')$ then $M',v' \models \varphi$. By slightly abusing the notation, and letting $(W,w)\Rel{b}^i (W,v)$ be equivalent to $w\Rel{b}^i v$, we can create the composed relation $(\llbracket i,(A,\laa),S\rrbracket \circ \Rel{b}^i)$. It then holds, by condition $(5)$, that $(M,w) (\llbracket i,(A,\laa),S\rrbracket \circ \Rel{b}^i) (M',v')$ implies $(M,w)\llbracket i, (A,\laa)\rrbracket (M',v')$. So, for all $M',v'$, if $(M,w) \llbracket i,(A,\laa)\rrbracket (M',v')$, then $M',v' \models \varphi$. So, $M,w \models \Pal{i,(A,\laa)}\varphi$. $\Box$

\section{Case Study and Mechanization}~\label{mech}

We apply the logic just developed to the formal analysis of the Air France 447 aviation incident. We also mechanize the formalization in the Coq Proof Assistant. Our mechanization follows similar work by Malikovi\'c and \v Cubrilo~\cite{delcoq1,delcoq2}, in which they mechanize an analysis of the game of Cluedo using Dynamic Epistemic Logic, based on van Ditmarsch's formalization of the game~\cite{ditmarsch}. It is commonly assumed that games must be adversarial, but this is not the case. Games need only involve situations in which players' payoffs depend on the actions of other players. Similarly, knowledge games need not be adversarial, and must only involve diverging information. Thus, it is appropriate to model aviation incidents as knowledge games of sorts, where players' payoffs depend on what others do, specifically the way the players communicate information with each other. The goal is to achieve an accurate situational awareness and provide flight control inputs appropriate for the situation. Failures to achieve this goal result in disaster, and often result from imperfect information flow. A formal model of information flow in these situations provides insight and allows for the application of formal methods to improve information flow during emergency situations.
\subsection{Air France 447}
This case study is based on the authoritative investigative report into Air France 447 performed and released by France's  Bureau d'Enqu\^etes et d'Analyses pour la S\'ecurit\'e de l'Aviation Civile (BEA), responsible for investigating civil aviation incidents and issuing factual findings\cite{airfrance}. The case is mechanized by instantiating, in Coq, the above logic to reflect the facts of the case. One challenge associated with this is that the readings about inputs present in aviation are often real values on a continuum, whereas for our purposes we require discrete values. We accomplish this by dividing the continuum associated with inputs and readings into discrete chunks, similar to how fuzzy logic maps defines predicates with real values\cite{fuzzy}.


This paper will formalize an excerpted instance from the beginning of the case, involving an initial inconsistency among airspeed indicators, and the subsequent dangerous input provided by the pilot. Formalized in the logic, the facts of the case allow us to infer that the pilot lacked negative introspection about the safety-critical data required for his action. This demonstrates that the logic allows information about the pilot's situational awareness to flow to the computer, via the pilot's actions. It likewise establishes a safety property to be enforced by the computer, namely that a pilot should maintain negative introspection about safety-critical data, and if he fails to do so, it should be re-established as quickly as possible.

According to the official report, at 2 hours and 10 minutes into the flight, a Pitot probe likely became clogged by ice, resulting in an inconsistency between airspeed indicators, and the autopilot disconnecting. This resulted in a change of mode from Normal Law to Alternate Law 2, in which certain stall and control protections ceased to exist. The pilot then made inappropriate control inputs, namely aggressive nose up commands, the only explanation for which is that he mistakenly believed that the aircraft was in Normal Law mode with protections in place to prevent a stall. This situation, and the inference regarding the pilot's mistaken belief, is modeled in the following application and mechanization of the logic.

\subsection{Mechanization in Coq}
The following mechanization demonstrates progress from the artificially simply toy examples normally analyzed in the literature to richer real-world examples. However, it does not represent the full richness of the approach. The actions and instrument readings mechanized in this paper are constrained to those most relevant to the case study. The approach is capable of capturing the full richness of all instrument reading configurations and actions available to a pilot. To do so, one needs to consult a flight safety manual and formally represent each action available to a pilot, and each potential instrument reading, according to the following scheme.

Before beginning, we note that our use of sets in the following Coq code requires the following argument passed to coqtop before executing: -impredicative-set. In CoqIDE, this can be done by selecting the `Tools' dropdown, then `Coqtop arguments'. Type in \emph{-impredicative-set}.

We first formalize the set of agents.
\begin{tcolorbox}
\begin{lstlisting}[language=Coq]
Inductive Agents: Set := Pilot | CoPilot | AutoPilot.
\end{lstlisting}
\end{tcolorbox}

Next we formalize the set of available inputs. These themselves are not actions, but represent atomic propositions true or false of a configuration.

\begin{tcolorbox}
	\begin{lstlisting}[language=Coq]
Inductive Inputs : Set := 
                   HardThrustPlus  | ThrustPlus 
                 | HardNoseUp      | NoseUp 
                 | HardWingLeft    | WingLeft
                 | HardThrustMinus | ThrustMinus
                 | HardNoseDown    | NoseDown 
                 | HardWingRight   | WingRight.
	\end{lstlisting}
\end{tcolorbox}

We represent readings by indicating which \emph{side} of the panel they are on. Typically, an instrument has a left-side version, a right-side version, and sometimes a middle version serving as backup. When one of these instruments conflicts with its siblings, the autopilot will disconnect and give control to the pilot.

\begin{tcolorbox}
\begin{lstlisting}[language=Coq]
Inductive Side : Set := Left | Middle | Right.
\end{lstlisting}	
	
\end{tcolorbox}

We divide the main instruments into chunks of values they can take, in order to provide them with a discrete representation in the logic. For example, the reading \emph{VertUp1} may represent a nose up reading between 0$\degree$ and 10$\degree$, while \emph{VertUp2} represents a reading between 11$\degree$ and 20$\degree$.

\begin{tcolorbox}
\begin{lstlisting}[language=Coq]
Inductive Readings (s : Side) : Set := 
          VertUp1 | VertUp2 | VertUp3 | VertUp4 
        | VertDown1 | VertDown2 | VertDown3 | VertDown4 
        | VertLevel | HorLeft1 | HorLeft2 | HorLeft3 
        | HorRight1 | HorRight2 | HorRight3 | HorLevel
        | AirspeedFast1 | AirspeedFast2 | AirspeedFast3 
        | AirspeedSlow1 | AirspeedSlow2 | AirspeedSlow3 
        | AirspeedCruise| AltCruise | AltClimb | AltDesc | AltLand.
\end{lstlisting}	
	
\end{tcolorbox}

We define a set of potential modes the aircraft can be in.

\begin{tcolorbox}
	\begin{lstlisting}[language=Coq]
Inductive Mode : Set := Normal | Alternate1 | Alternate2.
	\end{lstlisting}
\end{tcolorbox}

We define a set of global instrument readings representing the mode and all of the instrument readings, left, right, and middle, combined together. This represents the configuration of the instumentation.

\begin{tcolorbox}
\begin{lstlisting}[language=Coq]
Inductive GlobalReadings : Set := Global (m: Mode) 
                                            (rl : Readings Left) 
                                            (rm : Readings Middle) 
                                            (rr : Readings Right). 
\end{lstlisting}
\end{tcolorbox}

The set of atomic propositions we are concerned with are those representing facts about the instrumentation.

\begin{tcolorbox}
\begin{lstlisting}[language=Coq]
Inductive Atoms : Set := 
         | M (m : Mode)
         | Input (a : Inputs) 
         | InstrumentL (r : Readings Left) 
         | InstrumentM (r : Readings Middle) 
         | InstrumentR (r : Readings Right)
         | InstrumentsG (g : GlobalReadings).
\end{lstlisting}
\end{tcolorbox}

Next we follow Malikovi\'c and \v Cubrilo~\cite{delcoq1,delcoq2} in defining a set \emph{prop} of propositions in predicate calculus, distinct from Coq's built in type \emph{Prop}. The definition provides constructors for atomic propositions consisting of particular instrument reading predicate statements, implications, propositions beginning with a knowledge modality, and those beginning with a belief modality. Interestingly, modal logic cannot be directly represented in Coq's framework~\cite{lescanne}. We first define propositions in first-order logic, which we then use to define DASL. This appears to be the standard technique for mechanizing modal logics in Coq.

\begin{tcolorbox}
\begin{lstlisting}[language=Coq]	
Inductive prop : Set :=
| atm : Atoms -> prop
| imp: prop -> prop -> prop
| Forall : forall (A : Set), (A -> prop) -> prop
| K : Agents -> prop -> prop
| B : Agents -> prop -> prop
| Ck : list Agents -> prop -> prop
| Cb : list Agents -> prop -> prop.
\end{lstlisting}
\end{tcolorbox}

We use the following notation for implication and universal quantification.

\begin{tcolorbox}
\begin{lstlisting}[language=Coq]
Infix "==>" := imp (right associativity, at level 85).
Notation "\-/ p" := (Forall _ p) (at level 70, right associativity).
\end{lstlisting}
\end{tcolorbox}

We likewise follow Malikovi\'c and \v Cubrilo~\cite{delcoq1,delcoq2} by defining an inductive type \emph{theorem} representing a theorem of DASL. The constructors correspond to the Hilbert system, either as characteristic axioms, or inference rules. The first three represent axioms for propositional logic, then the rule Modus Ponens, then the axioms for the epistemic operator plus its Necessitation rule, then the doxastic operator and its Necessitation rule. Do not confuse the Necessitation rules with material implication in the object language. The final constructors capture the axioms relating belief and knowledge. The axioms for dynamic modal operators are defined separately, and are not included here.

\begin{tcolorbox}
\begin{lstlisting}[language=Coq]
Inductive theorem : prop -> Prop :=
   |Hilbert_K: forall p q : prop, theorem (p ==> q ==> p)
   |Hilbert_S: forall p q r : prop, 
                theorem ((p==>q==>r)==>(p==>q)==>(p==>r))
   |Classic_NOTNOT : forall p : prop, theorem ((NOT (NOT p)) ==> p)
   |MP : forall p q : prop, theorem (p ==> q) -> theorem p -> theorem q
   |K_Nec : forall (a : Agents) (p : prop), theorem p -> theorem (K a p)
   |K_K : forall (a : Agents) (p q : prop), 
          theorem (K a p ==> K a (p ==> q) ==> K a q)
   |K_T : forall (a : Agents) (p : prop), theorem (K a p ==> p)
   |B_Nec : forall (a : Agents) (p : prop), theorem p -> theorem (B a p)
   |B_K : forall (a : Agents) (p q : prop), 
          theorem (B a p ==> B a (p ==> q) ==> B a q)
   |B_Serial : forall (a : Agents) (p : prop), 
               theorem (B a p ==> NOT (B a (NOT p)))
   |B_4 : forall (a : Agents) (p : prop), theorem (B a p ==> B a (B a p))
   |B_5 : forall (a : Agents) (p : prop), 
          theorem (NOT (B a p) ==> B a (NOT (B a p)))
   |K_B : forall (a : Agents) (p : prop), theorem (K a p ==> B a p)
   |B_BK : forall (a : Agents) (p : prop), theorem (B a p ==> B a (K a p)).
  
\end{lstlisting}\end{tcolorbox}

We use the following notation for \emph{theorem}:

\begin{tcolorbox}	\begin{lstlisting}[language=Coq]
Notation "|-- p" := (theorem p) (at level 80).
\end{lstlisting}
\end{tcolorbox}

We encode actions as records in Coq, recording the acting pilot, the observability of the action (whether it is observed by other agents or not), the input provided by the pilot, and the preconditions for the action and the safety preconditions for the action, both represented as global atoms.

\begin{tcolorbox}
\begin{lstlisting}[language=Coq]
Record Action : Set := act {Ai : Agents; Aj : Agents; pi : PI; 
                                input : Inputs; c : GlobalReadings; 
                                c_s : GlobalReadings}.
\end{lstlisting}
\end{tcolorbox}

The variable \emph{c} holds the configuration representing the precondition for the action, while the variable \emph{c\_s} holds the configuration for the safety precondition.

	
We encode the precondition and safety precondition functions as follows.

\begin{tcolorbox}
\begin{lstlisting}[language=Coq]
Function pre (a:Action) : prop := atm (InstrumentsG (c a)).

Function pre_s (a : Action) : prop := atm (InstrumentsG (c_s a)).

\end{lstlisting}
\end{tcolorbox}

In the object language, the dynamic modalities of action and safe action are encoded as follows.

\begin{tcolorbox}\begin{lstlisting}[language=Coq]
Parameter aft_ex_act : Action -> prop -> prop.
Parameter aft_ex_act_s : Action -> prop -> prop.

\end{lstlisting}
\end{tcolorbox}

Many standard properties of logic, like the simplification of conjunctions, hypothetical syllogism, and contraposition, are encoded as Coq axioms. As an example, here is how we encode simplifying a conjunction into just its left conjunct.

\begin{tcolorbox}
\begin{lstlisting}[language=Coq]
Axiom simplifyL : forall p1 p2,
       |-- p1 & p2 -> |-- p1.
\end{lstlisting}
\end{tcolorbox}

We formalize the configuration of the instruments at 2 hour 10 minutes into the flight as follows.

\begin{tcolorbox}\begin{lstlisting}[language=Coq]
Definition Config_1 :=  (atm (M Alternate2)) & 
                         (atm (InstrumentL (AirspeedSlow3 Left))) & 
                         (atm (InstrumentM (AirspeedSlow3 Middle))) & 
                         (atm (InstrumentR (AirspeedCruise Right))).
\end{lstlisting}\end{tcolorbox}
The mode is Alternate Law 2, and the left and central backup instruments falsely indicate that the airspeed is very slow, while the right side was not recorded, but because there was a conflict, we assume it remained correctly indicating a cruising airspeed.

The pilot's dangerous input, a hard nose up command, is encoded as follows.

\begin{tcolorbox}\begin{lstlisting}[language=Coq]
Definition Input1 := act Pilot Pilot Pri HardNoseUp 
                       (Global Alternate2 (AirspeedSlow3 Left) 
                                            (AirspeedSlow3 Middle) 
                                            (AirspeedCruise Right))
                       (Global Normal (AirspeedCruise Left) 
                                       (AirspeedCruise Middle) 
                                       (AirspeedCruise Right)).
\end{lstlisting}\end{tcolorbox}

The action is represented in the object language by taking the dual of the dynamic modality, $\tlnot \Pal{i,(A,\laa)}\tlnot True$, equivalently $\PalPos{i}{(A,\laa)}True$, indicating that the precondition is satisfied and the action token is executed.

\begin{tcolorbox}\begin{lstlisting}[language=Coq]
Definition Act_1 :=  NOT (aft_ex_act Input1 (NOT TRUE)).
\end{lstlisting}
\end{tcolorbox}

The actual configuration satisfies the precondition for the action, but it is inconsistent with the safety precondition. The safety precondition for the action indicates that the mode should be Normal and the readings should consistently indicate cruising airspeed. However, in Config\_1, the conditions do not hold. Thus, the action is unsafe. From the configuration and the action, DASL allows us to deduce that the pilot lacks negative introspection of the action's safety preconditions.

Negative introspection is an agent's awareness of the current unknowns. To lack it is to be unaware of one's unknown variables, so lacking negative introspection about one's safety preconditions is to be unaware that they are unknown.

\begin{tcolorbox}\begin{lstlisting}[language=Coq]
Theorem NegIntroFailMode : 
                   |-- (Config_1 ==> 
                         Act_1 ==>
                         ((NOT (K Pilot (pre_s(Action1)))) &
                          (NOT (K Pilot (NOT (K Pilot (pre_s(Action1)))))))).
\end{lstlisting}\end{tcolorbox}

In fact, in general it holds that if the safety preconditions for an action are false, and the pilot executes that action, then the pilot lacks negative introspection of those conditions. We have proven both the above theorem, and the more general theorem, in Coq.

\begin{tcolorbox}\begin{lstlisting}[language=Coq]
Theorem neg_intro_failure : 
forall (A Ao : Agents) (pi : PI) (inp : Inputs) 
     (m : Mode) 
     (rl : Readings Left) (rm : Readings Middle) (rr : Readings Right) 
     (ms : Mode) 
     (rls : Readings Left) (rms : Readings Middle) (rrs : Readings Right) 
     phi,
|--  (NOT 
       (aft_ex_act 
          (act A Ao pi inp (Global m rl rm rr) (Global ms rls rms rrs)) 
       (NOT phi)) ==>
      NOT (atm (InstrumentsG (Global ms rls rms rrs))) ==>
        (NOT (K A (atm (InstrumentsG (Global ms rls rms rrs)))) & 
        (NOT (K A (NOT (K A (atm (InstrumentsG (Global ms rls rms rrs))))))))).

\end{lstlisting}
\end{tcolorbox}

This indicates that negative introspection about safety preconditions is a desirable safety property to maintain, consistent with the official report's criticism that the Airbus cockpit system did not clearly display the safety critical information. The logic described in this research accurately models the report's findings that the pilot's lack of awareness about safety-critical information played a key role in his decision to provide unsafe inputs. Furthermore, the logic supports efforts to automatically infer which safety-critical information the pilot is unaware of and effectively display it to him. 

	\section{Additional Case Studies}
	\noindent
	To illustrate the flexibility of this approach, we now formalize additional case studies in the logic. We begin with two additional aviation incidents, Copa Airlines flight 201 and Asiana Airlines flight 214, and then turn our attention to the domains of emergency health care and powerplant operations.
	
	We make use of the following lemma.
	
	\begin{eqnarray}
	\Bels{i} \varphi \iimplies \tlnot \Kns{i}\tlnot \varphi.
	\end{eqnarray}
	$\mathbf{Proof}$. From the fact that the belief modality is serial, it holds that
	\begin{eqnarray*} \Bels{i}\varphi \iimplies \BPoss{i} \varphi,\end{eqnarray*} which is equivalent to \begin{eqnarray*} \Bels{i} \varphi \iimplies \tlnot \Bels{i}\tlnot \varphi. \end{eqnarray*} Due to axiom EP1, it follows that \begin{eqnarray*} \Bels{i} \varphi \iimplies \tlnot \Kns{i}\tlnot \varphi. \end{eqnarray*} $\Box$
	%
	\subsection{Copa 201 and Asiana 214}
	\noindent
	Copa flight 201 departed Panama City, Panama for Cali, Colombia in June, 1992. Due to faulty wiring in the captain's Attitude Indicator, he incorrectly believed he was in a left bank position. In response to this, he directed the plane into an 80 degree roll to the right, which caused the plane to enter a steep dive. A correctly functioning backup indicator was available to the captain, and investigators believe that the captain intended to direct the backup indicator's readings to his own, but due to an outdated training module, the flip he switched actually sent his own faulty readings to the co-pilot's indicator. Approximately 29 minutes after takeoff, the plane crashed into the jungle and all passengers and crew perished. We formalize the moment at which the pilot provides the hard right roll input.
	
	\begin{enumerate}
		\item $\mathit{(Left AI HorLeft2)} \tland \tlnot\mathit{(Middle AI HorLeft2)} \dots$-------------------- configuration.
		\item $\PalPos{\mathit{pilot}}{\mathit{hardwingright}}\mathit{true}$------------------------------------------------ pilot input.
		\item $\Bels{\mathit{pilot}}(\mathit{Middle AI HorLeft2})$------------------------------------ from axiom PR, $\mathit{pre_s}$.
		\item $\tlnot \Kns{\mathit{pilot}}(\mathit{Middle AI HorLeft2})$------------------------------- from axiom K-Reflexive.
		\item $\Bels{\mathit{pilot}}\Kns{\mathit{pilot}}(\mathit{Middle AI HorLeft2})$--------------------------- from (3), axiom EP2.
		\item $\tlnot \Kns{\mathit{pilot}} \tlnot \Kns{\mathit{pilot}}(\mathit{Middle AI HorLeft2})$--------------------- from (5), B-Serial, EP1.
		\item $\tlnot \Kns{\mathit{pilot}}(\mathit{Middle AI HorLeft2}) \tland \tlnot \Kns{\mathit{pilot}} \tlnot \Kns{\mathit{pilot}}(\mathit{Middle AI HorLeft2})$-- (4), (6).
		\item $\tlnot(\tlnot \Kns{\mathit{pilot}}(\mathit{Middle AI HorLeft2}) \iimplies \Kns{\mathit{pilot}} \tlnot \Kns{\mathit{pilot}}(\mathit{Middle AI HorLeft2}))$ - (7).
	\end{enumerate}
	
	Asiana flight 214 from South Korea to San Francisco departed in the evening of July 6, 2013 and was schedule to land just before noon that morning. The weather was good and air traffic control cleared the pilots to perform a visual approach to the runway. The plane came in short and crashed against an embankment in front of the runway, resulting in the deaths of three passengers and 187 injured. The National Transportation Safety Board (NTSB) investigation found that the captain had mismanaged the approach and monitoring of the airspeed, resulting in the plane being too high for a landing. Upon noticing this, the captain selected a flight mode (flight level change speed) which unexpectedly cause the plane to climb higher. In response to this, the captain disconnected the autopilot and pulled back on the thrust. This caused an autothrottle (A/T) protection to turn off, so when the captain pitched the nose down, the plane descended fast than was safe, causing it to come down too quickly and collide with the embankment in front of the runway. We will formalize the moment at which the pilot pitches the nose down.
	
	\begin{enumerate}
		\item $\mathit{(A/T=Off)} \tland \mathit{(AirspeedSlow3)} \dots$----------------------------- configuration.
		\item $\PalPos{\mathit{pilot}}{\mathit{hardthrustminus}}\mathit{true}$------------------------------------------- pilot input.
		\item $\Bels{\mathit{pilot}}(\mathit{A/T=On})$------------------------------------------ from axiom PR, $\mathit{pre_s}$.
		\item $\tlnot \Kns{\mathit{pilot}}(\mathit{A/T=On})$------------------------------------- from axiom K-Reflexive.
		\item $\Bels{\mathit{pilot}}\Kns{\mathit{pilot}}(\mathit{A/T=On})$----------------------------------- from (3), axiom EP2.
		\item $\tlnot \Kns{\mathit{pilot}} \tlnot \Kns{\mathit{pilot}}(\mathit{A/T=On})$--------------------------- from (5), B-Serial, EP1.
		\item $\tlnot \Kns{\mathit{pilot}}(\mathit{A/T=On}) \tland \tlnot \Kns{\mathit{pilot}} \tlnot \Kns{\mathit{pilot}}(\mathit{A/T=On})$---------------- (4), (6).
		\item $\tlnot(\tlnot \Kns{\mathit{pilot}}(\mathit{A/T=On}) \iimplies \Kns{\mathit{pilot}} \tlnot \Kns{\mathit{pilot}}(\mathit{A/T=On}))$ ------------------ (7).
	\end{enumerate}
	
	The above formalizations follow the same format as that of Air France 447. A pilot provides an input whose safety precondition conflicts with one of the instruments in the configuration, and we infer that the pilot lacks negative introspection of the safety precondition. This is distinct from but related to the property that the pilots, in engaging in an unsafe action, are unaware of the unsafe instrument readings. Thus, we describe our first safety properties.
	\begin{enumerate}
		\item Safety Negative Introspection (SNI). If a safety precondition does not hold, then agent knows that he does not know it to hold. \begin{eqnarray*}
			\tlnot \mathit{pre_s}(\laa) \iimplies \Kns{i} \tlnot \Kns{i}\mathit{pre_s}(\laa)
		\end{eqnarray*}
		\item Unsafety Delivery (UD). If a safety precondition is false, then agent knows that it is false. \begin{eqnarray*}
			\tlnot \mathit{pre_s}(\laa) \iimplies \Kns{i}\tlnot \mathit{pre_s}(\laa)
		\end{eqnarray*}
	\end{enumerate}
	
	Our above formalizations show that SNI is false when a pilot provides an unsafe input. 
	Notice that UD implies SNI.
	
	$\mathbf{Proof}$. It suffices to show that $\Kns{i}\tlnot\varphi \iimplies \Kns{i}\tlnot\Kns{i} \varphi.$
	Assume $\Kns{i}\tlnot\varphi$ holds. From EP1, it follows that $\Bels{i}\tlnot\varphi$, and because knowledge is a normal modality, it follows that $\Kns{i}\Bels{i}\tlnot\varphi$ holds. From lemma 1, and again the fact that knowledge is normal modality, it follows that $\Kns{i}\tlnot\Kns{i}\tlnot\tlnot\varphi$, or equivalently, that $\Kns{i}\tlnot\Kns{i}\varphi$. $\Box$
	
	However, the converse does not hold. We can satisfy SNI when the safety precondition is false, the agent knows that he doesn't know it, but doesn't know that it is false. A counterexample consists of a model with three worlds: \{u,v\}. Let $\varphi$ be the safety precondition, with the following truth assignment: \{False, True\}. Let the epistemic relation include (u,v), (v,u), and the reflexive relations. Then at world u $\varphi$ is false, and $\Kns{i}\tlnot\Kns{i}\varphi$ is true, but $\Kns{i}\tlnot \varphi$ is false. 
	
	The formalization show that from the pilot's unsafe action, it follows that he lacks negative introspection of the safety precondition. 
	\begin{equation}
	\PalPos{i}{\laa}\mathit{true} \tland \tlnot \mathit{pre_s}(\laa) \iimplies \tlnot \Kns{i}\tlnot \Kns{i}\mathit{pre_s}(\laa)
	\end{equation}This situation violates SNI, because the pilot doesn't know that he doesn't know the safety precondition. Since UD implies SNI, UD is also violated.
	\begin{equation}
	\PalPos{i}{\laa}\mathit{true} \tland \tlnot \mathit{pre_s}(\laa) \iimplies \tlnot \Kns{i}\tlnot \mathit{pre_s}(\laa)
	\end{equation} So, from an unsafe action, we can also infer that the pilot does not know that the safety precondition is false, a stronger conclusion.  
	
	Thus, by restoring knowledge that the safety precondition is false, it follows that either the safety precondition is true, or the unsafe action is not executed. 
	\begin{eqnarray}
	\Kns{i}\tlnot \mathit{pre_s}(\laa) \iimplies \tlnot \PalPos{i}{\laa}\mathit{true} \tlor \mathit{pre_s}(\laa)
	\end{eqnarray}
	The pilot's knowledge in the antecedent implies that the safety precondition is false, so this simplifies to:
	\begin{equation}
	\Kns{i}\tlnot \mathit{pre_s}(\laa) \iimplies \tlnot \PalPos{i}{\laa}\mathit{true}.
	\end{equation}
	
	This squares with the standard game theoretic inference, wherein a rational agent with knowledge of the situation executes a good action. Because our model of knowledge and rationality is weaker, we make the weaker claim that a minimally rational pilot with knowledge of the safety-critical information does not execute a bad action.
\section{Future Work}~\label{futurework}
\noindent
The case study presented in this paper is overly simplified due to space constraints. Future work will 
undertake the task of extending the approach to other actions in the Air France 447 incident, and the safety-critical information expressed by them. For example, when both pilots provided conflicting inputs to the aircraft, the computer could have inferred that neither was aware of the other's actions. This will illustrate the use of the approach in a multi-agent context. Similarly, as recommended by an anonymous reviewer, we shall apply the approach to other aviation mishaps involving complicated safety-critical information flow, specifically Asiana Airlines Flight 214~\cite{asiana}.

An important extension of the foundational work provided by this paper is the construction of a system that takes advantage of the logic as a runtime safety monitor. It will monitor the pilot's control inputs and current flight configurations, and in the event that an action's safety preconditions do not hold, infer which instrument readings the pilot is unaware of and act to correct this. In order to avoid further information overload, the corrective action taken by the computer should be to temporarily remove or dim the non-safety-critical information from competition for the pilot's attention, until the pilot's unsafe control inputs are corrected, indicating awareness of the safety-critical information. Construction of a prototype of this system is underway.

\section{Conclusion}~\label{conc}
This paper has described Dynamic Agent Safety Logic (DASL), a logic for reasoning about safety-critical information flow. It formalized actions and knowledge in the way common to Dynamic Epistemic Logic, but also formalized the notion of safe actions and beliefs. Additionally, it formalized a more realistic model of human reasoning, capturing a weaker notion of knowledge than most epistemic logics, and modeled the logical relationship between knowledge and belief. It formalized a realistic notion of rationality. The logic was mechanized in the Coq proof assistant and applied to the case of Air France 447 to validate its usefulness as a formal method for aviation safety.

$\mathbf{Acknowledgements.}$ Seth Ahrenbach was partially supported by  NSF CNS 1553548. The author is grateful for the criticism and suggestions provided by anonymous reviewers, and for the very generous assistance from Alwyn Goodloe, Rohit Chadha, and Chris Hathhorn.

\end{document}